\documentclass[11pt, executivepaper]{article}
\usepackage[utf8]{inputenc}
\usepackage[T1]{fontenc}
\usepackage{natbib}
\usepackage{amsmath}
\usepackage{mathtools}
\usepackage{xcolor}
\usepackage{amsfonts}
\usepackage{systeme}
\usepackage{graphicx}
\usepackage{enumitem}
\usepackage{geometry}
\usepackage{hyperref}
\hypersetup{colorlinks= true, allcolors=blue}
\setcitestyle{aysep={}}
\begin{document}

\title{\textbf{The Relational Dissolution of the Quantum Measurement Problems}}

\author{Andrea Oldofredi\thanks{Contact Information: University of Lisbon, Centre of Philosophy, 1600-023 Lisbon, Portugal. E-mail: aoldofredi@letras.ulisboa.pt}}

\maketitle

\begin{abstract}
The Quantum Measurement Problem is arguably one of the most debated issues in the philosophy of Quantum Mechanics, since it represents not only a technical difficulty for the standard formulation of the theory, but also a source of interpretational disputes concerning the meaning of the quantum postulates. Another conundrum intimately connected with the QMP is the Wigner friend paradox, a thought experiment underlining the incoherence between the two dynamical laws governing the behavior of quantum systems, i.e\ the Schr\"odinger equation and the projection rule. Thus, every alternative interpretation aiming to be considered a sound formulation of QM must provide an explanation to these puzzles associated with quantum measurements. It is the aim of the present essay to discuss them in the context of Relational Quantum Mechanics. In fact, it is shown here how this interpretative framework dissolves the QMP. More precisely, two variants of this issue are considered: on the one hand, I focus on the ``the problem of outcomes'' contained in \cite{Maudlin:1995aa}---in which the projection postulate is not mentioned---on the other hand, I take into account Rovelli's reformulation of this problem proposed in \cite{Rovelli:2022}, where the tension between the Schr\"odinger equation and the stochastic nature of the collapse rule is explicitly considered. Moreover, the relational explanation to the Wigner's friend paradox is reviewed, taking also into account some interesting objections \emph{contra} Rovelli's theory contained in \cite{Laudisa:2019}. I contend that answering these critical remarks leads to an improvement of our understanding of RQM. Finally, a possible objection against the relational solution to the QMP is presented and addressed. 
\vspace{5mm}

\end{abstract}
\vspace{5mm}

\begin{center}
\emph{Accepted for publication in Foundations of Physics}
\end{center}

\clearpage 
\tableofcontents
\vspace{10mm}

\section{Introduction}
\label{intro}

The Quantum Measurement Problem (QMP) is arguably one of the most debated and discussed issues in the philosophy of Quantum Mechanics (QM), since it represents not only a technical difficulty for the standard formulation of the theory, but also a source of interpretational disputes concerning the meaning of the quantum postulates. According to many scholars this is motivated by the presence of physically ill-defined notions within the axioms of QM, as for instance ``measurement'' and ``observer''. More precisely, several authors argued that standard quantum theory does not provide a physical and causal explanation for the realizations of macroscopic measurement outcomes, remaining \emph{de facto} silent about what kind of dynamical processes are responsible for the actualization of particular experimental results (cf.\ \cite{Bell:2004aa}, \cite{Bricmont:2016aa}, \cite{Maudlin:1995aa} for more details). Other authors claim that the QMP originates from the incoherence between the two dynamical laws governing the behavior of quantum systems, i.e\ the Schr\"odinger equation and the projection rule (cf.\ \cite{Rovelli:2022}). The latter issue has been notoriously investigated by Wigner, who proposed a thought experiment highlighting the undesired consequences generated by this inconsistency (cf.\ \cite{Wigner:1995}). 

Since the QMP is generally considered one of the major conundra of quantum mechanics, and the Wigner friend paradox underlines a crucial defect of the theory, every alternative interpretation aiming to be considered a sound formulation of QM must provide an explanation to these problematic issues associated with quantum measurements. In fact, in the literature concerning the foundations of quantum theory, many essays have been written to show how well-known interpretations of QM solve the QMP. Notable examples are provided by \cite{Bohm:1952ab}, \cite{Holland:1993}, \cite{Durr:2009fk} that explain in great detail the resolution of the QMP offered by different approaches to the pilot-wave theory, \cite{Ghirardi:2016}, \cite{Bassi:2003} who discussed how macroscopic measurement outcomes are obtained from the Ghirardi-Rimini-Weber dynamics, or \cite{Everett:1957}, \cite{Wallace:2012aa} that show how the relative state formulation of QM and the many-world interpretation respectively face the measurement problem. 

In the present essay I will discuss both the QMP---or better, different formulations of the problem---and the Wigner friend scenario in the context of Relational Quantum Mechanics (RQM), an interpretation of the standard quantum formalism provided by Carlo Rovelli in the mid-nineties.\footnote{In \cite{Rovelli:1996} RQM is introduced for the first time; in this paper I assume that the reader is familiar with its physical and metaphysical principles. For space reasons I will not discuss them here; however, excellent introductions and presentations of this interpretation are contained in \cite{Rovelli:2016, Rovelli:2018, Rovelli:2022} and \cite{Laudisa:2019b}.} This framework finds its motivations in ``the experimental evidence at the basis of quantum mechanics'' that ``forces us to accept that distinct observers give different descriptions of the same events'' (\cite{Rovelli:1996}, p.\ 1638). Such a claim, which entails a novel perspective on the interpretation of QM, finds its roots in Einstein's idea behind special relativity, i.e.\ that physical observations depends on the specific reference frame in which these interactions occur.\footnote{Analyzing Rovelli's papers on RQM it clearly emerges how Einstein's relativity theories and their axioms influenced the relational interpretation of QM. For an interesting discussion on this topic the reader may refer to  \cite{Pienaar:2021} and \cite{Rovelli:2021}.} Furthermore, these insights on the meaning of non-relativistic quantum theory are also motivated by Rovelli's reflections on quantum gravity and general relativity---according to him, in fact, in these theories the relational features of spacetime clearly emerge---and from the ubiquitous presence of relational quantities in physics---from classical mechanics and electromagnetism to quantum field theory and cosmology (cf.\ \cite{Rovelli:2014}). 

The main novelty of Rovelli's approach is to take seriously into account the relational character of physics, applying it to the formalism of standard QM taken at face value. This strategy entails remarkable consequences not only for the worldview that RQM provides, but also for the way in which many conceptual issues affecting quantum theory are solved---as we will see in the following sections. According to this interpretative framework, in fact, the notion of observer-independent state of physical systems is abolished in favor of an observer-dependent description of reality. In RQM different observers may provide distinct characterizations of the same sequence of physical events, all of which are equally correct and non-contradictory. As a consequence, the values of measurable properties are \emph{relative to} specific observers. More generally, the relational character of Rovelli's view implies that the notion of an absolute reality should be abandoned:
\begin{quote}
a quantum mechanical description of a certain system (state and/or values of physical quantities) cannot be taken as an ``absolute'' (observer-independent) description of reality, but rather as a formalization, or codification, of properties of a system \emph{relative} to a given observer. Quantum mechanics can therefore be viewed as a theory about the states of systems and values of physical quantities relative to other systems. [...]
Therefore, I maintain that in quantum mechanics, ``state'' as well as ``values of a variable''---or ``outcome of a measurement''---are relational notions in the same sense in which velocity is relational in classical mechanics'' (\cite{Rovelli:1996}, pp.\ 1648-1649).
\end{quote}

It is worth noting that in RQM the word ``observer'' refers to \emph{every} physical object having a particular definite state of motion. Thus, in RQM ``observers'' are not necessarily conscious beings: in this theoretical framework an electron, an air molecule as well as a Geiger counter can equally play the role of observers. Hence, conscious agents do not have a privileged status. Referring to this, \cite{Rovelli:2021} point out that one should not overestimate the relevance of this notion in relational QM: the theory is not primarily concerned with observers, it is about events, i.e.\ physical facts taking place in spacetime and relative to particular physical systems. Indeed, RQM supports an event ontology recalling Wittgenstein's idea for which the world is the totality of facts\footnote{For simplicity, in the present essay I will consider events and facts as synonyms. The subtle difference between these two notions in RQM is not relevant for the purposes of this work and can be overlooked.}, and not of things:
\begin{quote}
the ontology of RQM is a sparse (``flash'') ontology of relational quantum events, taken as primitive, and not derived from any ``underlying'' representation (\cite{Laudisa:2019b}).
\end{quote}

To this regard, recent analyses of the event ontology of RQM introduced the distinction between relative and stable facts. As we will see in the remainder of the present paper, this dichotomy will be relevant for our discussion because it provides a deeper understanding of how relational QM actually works in measurement situations.\footnote{For a new perspective about the event ontology of RQM the reader may refer to \cite{Adlam:2022}, where the authors introduce an additional interpretative postulate to the existing corpus of principles shaping the theory. As a consequence of it, the authors argue that observer-independent facts do exists although they are described by relational quantum states. Due to spatial constraints I cannot discuss here in detail this new interpretation of RQM, thus, it will not mentioned in the remainder of the paper.} 

Against this background, the aim of the present essay is twofold: in the first place, I show how this interpretative framework dissolves the QMP. More precisely, two variants of this issue are considered: on the one hand, I focus on the well-known ``problem of outcomes'' contained in \cite{Maudlin:1995aa}---in which the projection postulate is not mentioned---on the other hand, I take into account Rovelli's reformulation of this problem proposed in \cite{Rovelli:2022}, where the tension between the Schr\"odinger equation and the stochastic nature of the collapse rule is explicitly considered. While the former can be seen as a problem of quantum theory without the projection postulate, the latter may be interpreted as a conundrum implied by the obscure nature of the collapse process.\footnote{I thank the anonymous referee for pointing out this distinction.}
In the second place, I review how RQM explains the Wigner's friend paradox, taking also into account some interesting objections \emph{contra} Rovelli's theory contained in \cite{Laudisa:2019}. I contend that answering these critical remarks leads to an improvement of our understanding of RQM.\footnote{In this essay Laudisa critically discusses two open problems of relational QM: (i) the QMP and (ii) the issue of locality. Here I will be concerned only with the former; for a recent discussion of locality in RQM the reader may refer to \cite{Rovelli:2019}, \cite{Pienaar:2019}, \cite{Adlam:2022}.} 
\vspace{2mm}

The paper is organized as follows: the two formulations of the QMP and the solutions offered by RQM are presented in detail in Section \ref{QMP}. The Wigner's friend paradox and its relational explanation as well as Laudisa's criticisms are discussed in Section \ref{WF}. In Section \ref{disc} a possible objection against the relational solution to the QMP is presented and addressed. Finally, Section \ref{conc} concludes the essay.

\section{The Quantum Measurement Problems}
\label{QMP} 

The present section aims at explaining how RQM solves two formulations of the QMP; we will begin by taking into account the well-known ``problem of outcomes'' as formulated in \cite{Maudlin:1995aa}, then we will consider \cite{Rovelli:2022} analysis of the measurement problem, where the inconsistency between the Schr\"odinger dynamics and the collapse postulate is investigated.

\subsection{The Problem of Outcomes}

In order to present the first version of the QMP let us consider the standard example of a pre-measurement situation of the observable $S_z$ on a system $s$, where the latter is described by a wave function $\psi$ which is in a superposition of two $z-$spin states $|\uparrow\rangle_z, |\downarrow\rangle_z$: 

\begin{align}
\label{super}
\psi=a|\uparrow\rangle_z+b|\downarrow\rangle_z.
\end{align}

\noindent Here $|\uparrow\rangle_z$ corresponds to the $z-$spin-up state, $|\downarrow\rangle_z$ to the $z-$spin-down state, and the coefficients $|a|^2, |b|^2$ give the probability to find $s$ in the $z-$spin-up/down state respectively.
From QM, we assume that the wave function of a given system contains all information about its state, providing its \emph{complete} description. In this particular example $\psi$ represents a superposition of $z-$spin states, meaning that the superposition ``$z$-spin-up \emph{and} $z$-spin-down'' is a consistent description of the state in which a quantum particle may be (i.e.\ it is completely admitted by the theory's formalism as a proper physical state of a certain quantum object). This formal fact is usually taken to entail that prior to a spin measurement the particle has \emph{indefinite} $z-$spin, being neither in the $z$-spin-up state \emph{nor} in the $z$-spin-down state. 

In addition, before the measurement we assume that the macroscopic experimental device used to measure this quantity is in the ready state $\Phi_0$, i.e.\ in a state pointing to a neutral direction, whereas the other admissible pointer's positions will be $\Phi_1$ that is associated with the eigenvalue $+\hbar/2$, and $\Phi_2$ that is associated with $-\hbar/2$ respectively. Conforming to standard quantum theory we know that prior to the observation of $S_z$ the states of the system and the apparatus are independent, i.e.\ described by a product wave function. However, given the evolution provided by the Schr\"odinger equation we obtain the macroscopic superposition:
\begin{align}
\label{macrosuperpos}
(a|\uparrow\rangle_z+b|\downarrow\rangle_z)\Phi_0\longrightarrow a|\uparrow\rangle_z\otimes|\Phi_1\rangle+b|\downarrow\rangle_z\otimes|\Phi_2\rangle.
\end{align}

\noindent Thus, assuming (i) that wave functions provide a complete, observer-independent description of quantum systems, and (ii) that their dynamics is uniquely and completely described by the Schr\"odinger evolution, the microscopic superposition of states is amplified to the \emph{macroscopic scale}, as we can easily see from the r.h.s\ of equation \eqref{macrosuperpos}. Hence, it follows that also macroscopic objects can be in superpositions, contradicting empirical evidence which speaks in favor of the uniqueness and definiteness of measurement outcomes. This is in essence of the famous ``problem of outcomes'', a particular version of the measurement problem affecting quantum theory. Following \cite{Maudlin:1995aa}, it may be schematically stated as follows:
\begin{enumerate}
   \item[\emph{P.\ 1}]Wave functions provide a complete description of quantum systems;
   \item[\emph{P.\ 2}]Wave functions evolve according to the unitary evolution provided by the Schr\"odinger equation; 
   \item[\emph{P.\ 3}]Measurements have a unique determinate outcome.
\end{enumerate}

Remarkably, any pair of these propositions is consistent and entails the falsity of the third one, but the conjunction of the three statements generates inconsistencies with experimental evidence.\footnote{For a simple proof of this fact the reader should refer to \cite{Maudlin:1995aa}, pp.\ 7-8.} Interestingly, it is possible to understand how the various interpretations of QM solve the QMP by looking at what statement they reject. For instance, Bohmian mechanics denies the first proposition, since in every account of the pilot-wave theory the wave function does not exhaust the description of physical systems, which is supplemented by additional variables (particles' positions). Similarly, spontaneous collapse theories discard the second proposition, since they modify the Schr\"odinger equation and implement a non-linear dynamics, while Everett's relative state formulation of QM and the many worlds interpretation reject the last sentence, given that according to these frameworks every possible outcome is actualized in different branches of the wave function or in different worlds respectively. In the remainder of the section we will see how Rovelli's theory dissolves the problem of outcomes.

\subsection{The Relational Solution}

In order to understand how RQM deals with the QMP let us follow the usual strategy and see what propositions do not hold in the context of Rovelli's theory.  

In the first place, proposition \emph{P.\ 1} expresses the idea that the usual quantum mechanical wave function describes completely the inherent state of a quantum system, which is taken to be perspective-independent. Trivially, this is at odds with the basic tenets of RQM. However, such a statement can be retained with little effort also in the latter theory, since it is true that in RQM $\psi$ provides the maximal information available about physical systems relative to specific observers. Therefore, by taking into account the principle of relativization of states, we can rephrase Proposition 1 as follows:
\begin{quote} 
\emph{P.\ 1'} The quantum mechanical wave function $\psi$ describes completely the state of a system $s$ \emph{relative} to an observer $O$. 
\end{quote}
\noindent The reformulation of the first statement of the measurement problem entails that the same system may be described differently by distinct observers. More specifically, \emph{P.\ 1'} implies that distinct wave functions are employed by diverse observers in order to describe a certain physical system $s$; such descriptions will be however equally correct. In addition, according to RQM---in agreement with standard QM---the description of physical systems  provided by wave functions is complete, i.e.\ it does not require the introduction of additional parameters as done e.g.\ in hidden variables theories. Hence, RQM does not solve the measurement problem by rejecting \emph{P.\ 1}.
 
Our analysis gets more interesting by considering Proposition 2. To this regard, although the wave function in RQM dynamically evolves according to the Schr\"odinger equation, the latter does not exhaust the dynamics of the theory, since when interactions occur the unitary evolution is suppressed. Hence, the complete dynamics of RQM includes the projection postulated, exactly as in modern formulations of QM. Consequently, the above Proposition 2 is incorrect in the context of Rovelli's theory. To make an accurate statement concerning the physical content of the latter we should write:
\begin{quote}
\emph{P.\ 2'} Wave functions evolve according to Schr\"odinger evolution until a measurement is performed. When interactions occur the collapse rule suppresses the unitary dynamics. 
\end{quote}

It is worth stressing here that although RQM includes the projection postulate among its laws, the interpretation of such a rule in the relational context is completely different w.r.t.\ the usual one provided in standard QM. In fact, in RQM the suppression of the unitary dynamics is not caused by stochastic quantum jumps: since in Rovelli's theory $\psi$ is not considered a real object but rather a mere computational tool, nothing physical is literally collapsing in measurements interactions (cf.\ \cite{Rovelli:2016, Rovelli:2022}). Alternatively, given that $\psi$ ``stores'' the maximal information that a certain observer may have respect to a physical system to compute future predictions, the collapse of the wave function is interpreted as an information update relative to a certain agent concerning the value of some magnitude measured on a particular system.\footnote{The interpretation of the collapse postulate in RQM is similar to that provided by QBism. For discussion cf.\ \cite{Pienaar:2021a}.} The wave function, then, does not have any ontological significance in RQM, contrary to standard quantum theory.\footnote{The ontological status of the wave function in RQM has been discussed in several places, the reader may refer to \cite{Smerlak:2007}, \cite{Rovelli:2016}, \cite{Laudisa:2019b}, \cite{Oldofredi:2020}.}

In this respect, Rovelli's views are close to the Heisenberg picture, which is about the change of physical variables in time. Indeed, contrary to the Schr\"odinger picture, where the wave function describes completely the state of a system and evolves unitarily, in the Heisenberg picture $\psi$ just encodes information about past interactions, and it changes only as a result of another future interaction. Referring to this, Smerlak and Rovelli claimed that in RQM ``[w]hat evolves with time are the operators, whose expectation values code the time-dependent probabilities that can be computed on the basis of the past quantum events'' (\cite{Smerlak:2007}, p.\ 431).\footnote{As a consequence in RQM also the notion of ``wave function of the universe'' present in several interpretations of quantum theory---as for instance in Everett's relative state formulation of QM, the Many worlds interpretation, Bohmian mechanics, \emph{etc.}---is rejected.}

In addition to these remarks about the interpretation of $\psi$---which explain why the collapse postulate does not generate conceptual conundra---it is relevant to note that the exact details about the mechanisms causing the suspension of the unitary dynamics cannot be available in RQM. The reason is the following: if information is given in terms of correlation, a system cannot be correlated with itself, in turn exact information about the machinery which generates the suspension of the Schr\"odinger evolution are not accessible:
\begin{quote}
Since between times $t_1$ and $t_2$ the evolution of $s$ is affected by its interaction with $O$, the description of the unitary evolution of $s$ given by $O$ breaks down. The unitary evolution does not break down for mysterious
physical quantum jumps, due to unknown effects, but simply because $O$ is not giving a full dynamical description of the interaction. $O$ cannot have a full description of the interaction of $s$ with himself ($O$), because his information is correlation, and there is no meaning in being correlated with oneself
(\cite{Rovelli:1996}, pp.\ 1666-1667). 
\end{quote}
\noindent This fact should be simply considered as a descriptive limit of RQM, and not an ontological drawback affecting this framework (more on this issue in the next section).\footnote{Cf.\ \cite{Dorato:2016} for a good discussion of this issue.} Thus, it is the suppression of the unitary evolution via the collapse rule that allows RQM to overcome the problem of outcomes, however, in this theoretical framework such a postulate does not cause additional interpretational troubles.

Although we have seen that RQM rejects Proposition 2, it is interesting to note that also Proposition 3---which prescribes that measurements have a unique, observer-independent outcome---does not carry over within such a theory. In fact, \emph{P.\ 3} is true in RQM only \emph{relative to} an observer $O$ that measures a certain magnitude on a given system $s$. The agent $O$ will certainly obtain a unique, definite value via the interaction with the observed system. However, in virtue of the principle of relativity of states, another observer $W$ may ascribe a different state to the whole experimental situation, as we will see in the next section. Referring to this, however, it should be underlined that the requirement of internal consistency establishes that if $W$ interacts with the complex system formed by $s+O$---provided that $O$ does not destroy the information obtained measuring a certain observable on $s$---they will obtain the same result registered by $O$. On the other hand, without a direct interaction these two observers will in general assign different representations of the same sequence of physical events. Therefore, in the relational context the following is true: 
\begin{quote}
\emph{P.\ 3'} Measurements performed on a system $s$ have a unique determinate outcome relative to a specific observer $O$ that performed the observation. 
\end{quote}

Concerning the rejection of \emph{P.\ 3}, it is interesting to observe that RQM behaves differently from both the many-worlds interpretation as well as from Everett's original proposal, which are the main solutions to the QMP refuting Proposition 3. Contrary to the former, in a relational context there are no splitting of worlds; Rovelli's theory, in fact, is single-world interpretation of the quantum formalism.\footnote{Another crucial difference between these two interpretations is that in RQM the wave function of the universe, which play a central ontological role in MWI, does not exist.} More interesting is the difference existing between Everett's and Rovelli's perspectives. The first thing to highlight is that Everett aimed at solving the QMP by getting rid of the collapse postulate from the principles of QM, proposing a version of quantum theory that he defined ``pure wave mechanics''. According to this view, every possible outcome of a certain measurement is actualized in different branches of the wave function---which is the central object of the theory. Interestingly, each branch of $\psi$ corresponds to the relative state of a particular observer, which is itself part of the quantum mechanical description of physical situations. Referring to this, \cite{Barrett:2018} underlines that Everett ``believed that he could deduce the standard statistical predictions of quantum mechanics (...) in terms of the subjective experiences'' of such observers. 

As we have already seen, things are remarkably different in RQM where (i) \emph{not} every possible measurement result is actualized, (ii) wave functions do not split in different branches representing subjective experiences of some agents, and (iii) the projection postulate is retained. Taking into account the $z-$spin measurement employed to discuss Maudlin's formulation of the QMP, in fact, an observer $O$ performing the observation of the quantity $S_z$ on a system $s$ will obtain only one definite result. The relativization of states in Rovelli's theory claims that another observer $W$ may assign a different description of the same sequence of physical events, as we will see in detail in the next section.

Thus, RQM provides a different strategy to solve the problem of outcomes w.r.t.\ these well-known and established interpretations of QM.

\subsection{Unitarity vs. Collapse: Rovelli's Formulation of the QMP}

Before concluding our discussion, we have to take into consideration another approach to the QMP contained in \cite{Rovelli:2022}. According to Rovelli, the essence of the measurement problem lies in the alleged incoherence between the unitary evolution of the Schr\"odinger equation and the collapse postulate, he then reformulates the issue as follows.

It is well-known that the probability amplitude $W(b(t), a)$ for a certain fact $b$ to happen at time $t$ evolves according to the linear and unitary dynamics provided by the Schr\"odinger equation. On the other hand, the collapse rule affirms that the probability changes only when a measurement is performed, i.e.\ when a certain fact occurs. The contradiction between these two laws of QM arises, Rovelli contends, ``because of quantum interference: interference effects in the unitary evolution are cancelled by the projection'' (\cite{Rovelli:2022}, p.\ 1058).

In order to express this point more formally one can consider the following situation: suppose a particular fact $a$ happened, and one among possible facts $b_i$ (with $i=(1, \dots, N)$) may occur, then 
\begin{quote}
[b]y composition of probabilities, we expect the probability $P(c)$ for a further fact $c$ to happen to be given by
\begin{align*}
P_{collapse}(c|a)=\sum_i P(c|b_i)P(b_i|a),
\end{align*}
\noindent where $P(b|a)$ is the probability for $b$ to happen, given $a$. From the relation
between probability and amplitude
\begin{align}
\label{WW}
P_{collapse}(c|a)=\sum_i P|W(c, b_i)|^2|W(b_i, a)|^2.
\end{align}
But since quantum probabilities are squares of amplitudes and amplitudes sum,
linear evolution requires
\begin{align*}
P_{unitary}(c|a)=|W(c,a)|^2=|\sum_i W(c, b_i)W(b_i, a)|^2 \\
\neq \sum_i |W(c, b_i)|^2|W(b_i, a)|^2=P_{collapse}(c|a)
\end{align*}
(\emph{ibid.}, pp.\ 1058-1059).
\end{quote}

Thus, Rovelli concludes, the probability for the fact $c$ is different in the case one considers the unitary dynamics or the collapse rule, and this is in virtue of the quantum interference phenomena. Provided that we have already seen that the collapse postulate does not generate conceptual issues in RQM, we can still ask the following: what is the relational strategy to avoid this undesired inconsistency concerning quantum probabilities? The answer is very simple and has to be found in the labelling of facts (which will be explained in more detail in the next section). According to the principles of RQM, physical events happen relative to certain systems functioning as observers, so that generally \eqref{WW} does not hold in the case that facts $b_i$ and $c$ are labeled w.r.t.\ different agents, since the probability amplitudes $W(b,a)$ ``determine probabilities \emph{only} if $a$ and $b$ are relative to the same system''. 

Recalling the third person scenario from \cite{Rovelli:1996}---which is used by Rovelli in order to show how the relativization of states works---we can easily exemplify the above discussion. Suppose that the quantity $S_z$ is measured on the system $s$ by an agent $O$ inside their lab. The interaction between $s$ and $O$ causes the suppression of the unitary evolution---and therefore the suppression of interference effects---generating a unique, definite outcome for the system $s$ with respect to the observer $O$. Thus, the actualization of a particular value between $\pm\hbar/2$ is a fact relative to $O$, but not relative to another observer $W$ who is outside the room. Consequently, the probability for the actualization of a certain value for the magnitude $S_z$ w.r.t.\ $W$ will still involve interference effects. However, taking into account (i) the processes of wave function collapse and decoherence that occurred during the interaction between $s$ and $O$ and (ii) the system formed by $O$ and the environment in which the observation took place, the actualization of a specific experimental outcome can be considered a stable fact also for $W$. This entails that if agent $W$ interacts directly with the complex system $s+O$, $W$ will obtain results that are consistent with those found by the agent $O$. Thus, we can safely conclude this section claiming that RQM provides a detailed answer also to this formulation of the QMP. 

Interestingly, in the last paragraph we anticipated some details which are relevant in order to understand how RQM explains the Wigner's friend paradox to which we now turn.

\section{Relational QM and the Wigner's Friend Paradox}
\label{WF}

After having discussed how Rovelli's theory dissolves the QMP, in this section I am firstly going to show how this interpretative framework explains the Wigner's friend scenario, and secondly how it is possible to provide an answer to the interesting objections contained in \cite{Laudisa:2019} against RQM---showing that these criticisms do not threat its internal coherence.

The Wigner friend paradox is a prominent though experiment proposed by Eugene Wigner to show the inconsistency between the two fundamental laws of quantum theory, i.e.\ the linear, deterministic Schr\"odinger equation and the stochastic collapse postulate (cf.\ \cite{Wigner:1995}). Such an argument can be summarized in a few words as follows. 

Suppose that a friend of Wigner $O$ is performing a measurement of the $z-$spin on a quantum particle $s$ in their laboratory. Let's also assume that $s$ is not in an eigenstate of $S_z$, so that before the observation---at some arbitrary initial time $t_1$---the system is in the superposition encountered in the previous section:
\begin{align}
\label{suppos}
a|\uparrow\rangle_z+b|\downarrow\rangle_z.
\end{align}

\noindent Now, at a later time $t_2$ the agent $O$ performs the measurement of $S_z$ on $s$; consequently, according to the standard formulation of QM the state of the observed system collapses in one of the possible eigenstates of the measured quantity---for the sake of the argument, let's stipulate that $O$ finds the $z-$spin-up state $|\uparrow\rangle_z$ upon measurement. Hence, this sequence of physical events \textbf{E} relative to $O$ can be formalized as follows:

\begin{align}
\left.\begin{aligned}
\label{E}
t_1 \longrightarrow t_2\\
a|\uparrow\rangle_z+b|\downarrow\rangle_z \longrightarrow |\uparrow\rangle_z
\end{aligned}\right\} \textbf{E}
\end{align}

Now, Wigner ($W$) is outside the laboratory and describes the experimental situation into the room by using standard QM, knowing that $O$ will measure the $z-$spin of $s$. In this thought experiment $W$ is an external observer, and the lab should be considered for all practical purposes an isolated system. Consequently, from the linear dynamics of the Schr\"odinger equation it follows that according to $W$'s perspective the system $s+O$ is in a superposition of states:

\begin{equation}
\left.\begin{aligned}
\label{Estar}
t_1 \longrightarrow t_2\\
(a|\uparrow\rangle_z+b|\downarrow\rangle_z)\otimes |O-ready\rangle \longrightarrow a|\uparrow\rangle_z\otimes|O_\uparrow\rangle+b|\downarrow\rangle_z|O_\downarrow\rangle
\end{aligned}\right\} \textbf{E'}
\end{equation}

\noindent Remarkably, $W$ assigns a superposition state to the laboratory, although $O$ recorded a definite result after the $z-$spin measurement. It is only when $W$ asks directly to $O$ which outcome they obtained that the macroscopic superposition \textbf{E'} breaks. Nonetheless, well before $W$ asked his question, the state of the system $s$ was already definite, i.e.\ the superposition \eqref{suppos} was already decohered in the laboratory. Thus, Wigner concludes, the state after the interaction between $s$ and $O$ was already either $|\uparrow\rangle_z\otimes|O_\uparrow$ or $|\downarrow\rangle_z\otimes|O_\downarrow\rangle$, and not the one represented by the r.h.s.\ of \eqref{Estar}. Here is the contradiction in Wigner's words: 
\begin{quote}
the state described by the wave function $a(|\uparrow\rangle_z\otimes |O_\uparrow\rangle)+b(|\downarrow\rangle_z\otimes |O_\downarrow \rangle)$ describes a state that has properties which neither $|\uparrow\rangle_z\otimes|O_\uparrow\rangle$ or $|\downarrow\rangle_z\otimes|O_\downarrow\rangle$ has. [...] [T]he wave function $a(|\uparrow\rangle_z\otimes |O_\uparrow\rangle)+b(|\downarrow\rangle_z\otimes |O_\downarrow\rangle)$ (which also follows from the linearity of the equations) appears absurd because it implies that my friend was in a state of suspended animation before he answered my question (\cite{Wigner:1995}, p.\ 256, notation adapted). 
\end{quote}

Wigner's aim was to show that QM, in virtue of its incoherent laws, generates contradictory descriptions of the same reality. According to him, in order to resolve the tension existing between the dynamical postulates of quantum theory, and to make sense of them, the notion of consciousness, or of a conscious mind, should be introduced into the quantum picture in order to break the superposition of states described above. In Wigner's view, the mind---or more generally a being with consciousness---cannot be in a superposition of states, and thus it assumes a privileged status in quantum theory.\footnote{For further details on Wigner's view cf.\ \cite{Wigner:1995}.}

\subsection{The Relational Interpretation of the Paradox}

Wigner's argument generated intense discussions among the experts working on the foundations of quantum theory, and it is still vividly discussed to the present day.\footnote{Cf.\ \cite{Frauchiger:2018}, \cite{Bong:2020}, \cite{Brukner:2021}.} Indeed, it is well known how the most popular interpretations of QM solve this paradox and its new extensions (cf.\ for instance \cite{Lazarovici:2019}). Referring to this, it is worth noting that \cite{Rovelli:1996} explained how RQM works discussing a physical situation which is formally and logically equivalent to the Wigner friend scenario---what Rovelli called the ``Third Person Problem''. Let us then discuss the latter in order to understand the relational interpretation of Wigner's thought experiment. 

In the context of RQM the two sequences of events $\textbf{E}$ and $\textbf{E'}$ encountered above represent two different yet equally correct descriptions of the same physical situation. Following the principles of the theory, in fact, $\textbf{E}$ and $\textbf{E'}$ provide diverse characterizations of the experimental context under consideration, since they are relative to two distinct observers $O$ and $W$ respectively. According to the former, the system $s$ possesses a sharp value for $S_z$, whereas conforming to the latter (i) $s$ is \emph{not} in the state $|\uparrow\rangle_z$, and (ii) the measuring device does not indicate any sharp value of the $z-$spin. Indeed $W$ describes the complex system $s+O$ in a macroscopic superposition of states. 

Since RQM affirms that different observers can provide distinct descriptions of the same sequence of physical events, it follows that \emph{both} descriptions $\textbf{E}$ and $\textbf{E'}$ are equally valid and not contradictory. Thus, one concludes that the notion of ``state of a system'' must always be relative to another physical system. In particular, in relational QM a certain object has a definite value for a quantity \emph{only relative} to another observer. It follows, then, that the notions of ``quantum state'', ``measurement outcome'' and ``value of a variable'' are purely relational concepts in this context. Consequently, \emph{contra} Wigner, in RQM there is no need to introduce conscious agents to resolve the tension created by the different descriptions of events $\textbf{E}$ and $\textbf{E'}$, since they are both correct characterizations of a relational reality. This is, in essence, the message conveyed by the third person problem.\footnote{Three main consequences can be drawn from our discussion. Firstly, in RQM the separation between observed system and observer cannot be univocally determined, \emph{bona pace} Wigner---in our example agent $O$ observes $s$, but it is part of the composite observed system $s+O$ according to Wigner's perspective (more abut this issue in Section \ref{disc}). Secondly, as we already underlined, according to Rovelli's theory there is no privileged observer since all physical systems are equivalent: ``[n]othing a priori distinguishes macroscopic systems from quantum systems. If the observer $O$ can give a quantum description of the system $s$, then it is also legitimate for an observer $W$ to give a quantum description of the system formed by the observer $O$'' (\cite{Rovelli:1996}, p.\ 1644, notation adapted). Thirdly, RQM provides a \emph{complete} description of the world, because there is neither a deeper theory, nor hidden parameters describing how an absolute reality behaves, as we have already pointed out in the previous section studying the first proposition of the problem of outcomes.}

Interestingly, the relational solution of the Wigner friend paradox can be explained even more precisely with the aid of recent developments about Rovelli's theory contained in \cite{Rovelli:2021a} and \cite{Rovelli:2022}; particularly useful for us is the distinction between \emph{relative} and \emph{stable} facts.

According to Di Biagio and Rovelli relative facts can be generally defined as the interaction of a certain physical system---in our example $s$---with another system, in the case at hand $O$.\footnote{It is worth recalling that such interactions involve wave function collapses.} More precisely, if in virtue of the interaction occurred between $s$ and $O$, the variable representing the final state of $O$ depends on the value of $S_z$ measured on $s$, then such a value constitutes a fact \emph{relative} to the agent $O$, as \cite{Rovelli:2021a} claim (p.\ 3). These authors call $O$ a ``context'' for the actualization of a specific value of $S_z$ for $s$, thus, ``[t]he interaction with the context determines the fact that a certain variable has a value in that context'' (\emph{ibid.}). In this particular case, we can also say that the post-collapse actualization of a particular value for $s$ constitutes a \emph{stable} fact for $O$.\footnote{According to these authors stable facts are a proper subset of relative facts, i.e.\ all stable facts are also relative, but the converse does not hold: not every relative fact is stable.} The definition of stable facts provided by Di Biagio and Rovelli is the following: knowing that one among $N$ possible mutually exclusive facts $a_i$ with $i=(1, ..., N)$ occurred, the probability $P(b)$ for another fact $b$ to happen is given by
\begin{align}
\label{fact}
P(b)=\sum_i P(b|a_i)P(a_i),
\end{align}
\noindent where $P(a_i)$ represents the probability that $a_i$ has happened, and $P(b|a_i)$ is the conditional probability for the occurrence of the fact $b$ given $a_i$. In our example we are considering a fact about the system $s$ relative to the agent $O$, thus, fact $b$ corresponds to the pointer variable pointing in the $z-$spin-up direction, knowing that fact $a$ represents the post-collapse actualization of $|\uparrow\rangle_z$. Notably, these two facts are relative to $O$, consequently this formula holds:
\begin{align}
\label{fact2}
P(b^{(O)})=\sum_i P(b^{(O)}|a_i^{(O)})P(a_i^{(O)}).
\end{align}
\noindent We can therefore conclude that the actualization of a definite value for the magnitude $S_z$ measured on $s$ is a stable fact for $O$. In general, however, for any observer $W\neq O$ \eqref{fact2} is not valid; this in turn implies that
\begin{align}
\label{fact3}
P(b^{(W)})=\sum_i P(b^{(W)}|a_i^{(O)})P(a_i^{(O)})
\end{align}
\noindent does not always holds.\footnote{In those particular cases in which \eqref{fact3} holds for also for another observer $W$, then we can say that $a_i^{(O)}$ is stable also for $W$.} In the context of the Wigner friend scenario, the failure of the above equation for $W$ is due to the presence of interference effects: since the composite system $s+O$ can be considered for all practical purposes isolated for $W$, Wigner will represent the facts relative to the system $s+O$ according to \eqref{Estar}. Nonetheless, as Di Biagio and Rovelli underline, probabilities computed from that superposed state violate \eqref{fact3}, and therefore the actualization of the state $|\uparrow\rangle_z\otimes|O_\uparrow\rangle$ is not a stable fact for $W$: 

\begin{quote}
Equation \eqref{fact} holds only if $b$ and $a_i$ are facts relative to the same system, but fails in general if used for facts relative to different systems (\cite{Rovelli:2021a}, p.\ 3, equation number adapted).
\end{quote}

Remarkably, what makes a relative fact stable is quantum decoherence, an ubiquitous phenomenon that occurs as soon as a large number of quantum systems are involved. This is relevant for our discussion since via decoherence it is possible to show how the actualization of $|\uparrow\rangle_z$ relative to $O$ can be a stable fact also for Wigner. In order to achieve this result one has to consider two systems $O$ and $E$---where $E$ represents the environment in which the measurement of $S_z$ took place---and the variable corresponding to the possible final states of $O$, let us call it $L_O$ for simplicity. Let $O_{a_i}$ be the eigenvalues of $L_O$.  According to standard QM, the generic state of the complex system $O+E$ can be written as:
\begin{align*}
|\psi\rangle=\sum_i c_i|O_{a_i}\rangle\otimes|E_i\rangle,
\end{align*}
\noindent where $E_i$ represent the normalized states of the environment. Now, suppose that $W$ does not interact with $E$, as in the Wigner friend scenario. The probability of a possible fact relative to $W$ resulting \emph{from the interaction} between $W$ and $E$ can be computed via the density matrix obtained tracing over $E$:
\begin{align*}
\rho=\textrm{tr}_E |\psi\rangle\langle\psi|=\sum_i |c_i|^2|O_{a_i}\rangle \langle O_{a_i}|.
\end{align*}

\noindent Finally, establishing that $P(O_{a_i}^{(E)})=|c_i|^2$ we obtain

\begin{align*}
P(b^{(W)})=\sum_i P(b^{(W)}|O_{a_i}^{(E)})P(O_{a_i}^{(E)})
\end{align*}
\noindent meaning that the probabilities for facts relative to $W$ computed from the possible values of $L_O$ satisfy equation \eqref{fact3}. This result has remarkable implications: (i) the value of the variable $L_O$---corresponding to the final state of the agent $O$ after the observation of $S_z$---is \emph{stable for $W$}\footnote{Here I am simplifying the physical description of the actual situation since small interference effects should be taken into consideration. For a more detailed account of this discussion see \cite{Rovelli:2021a} Section 3.}, (ii) events concerning the systems $O$ and $E$ can generate a stable fact for another observer that has not yet interacted with them. Consequently, we can claim that also for $W$ the state of $O$ has effectively collapsed into one of the possible measurement outcomes (with probability $P(O_{a_i})=|c_i|^2$) although $W$ did not interact with $O$. 

An interesting question, then, is to ask how different perspectives may coexist without generating contradictions. We have already answered to this issue studying the second proposition of the problem of outcomes, however, let us give another, more specific reply. Recently, \cite{Rovelli:2021} stressed that one of the interpretative postulates of RQM is the following: ``[i]n the Wigner's friend scenario, if $W$ measures $s$ on the same basis on which $O$ did, then appropriately interacts with $O$ to `check the reading' of a pointer variable (i.e.\ by measuring $O$ in the appropriate `pointer basis'), the two values found are in agreement''.\footnote{A similar idea was already stated in \cite{Rovelli:1996} as follows: ``if $W$ knows that $O$ has measured $S_z$, and then she measures $S_z$, and then she measures what $O$ has obtained in measuring $S_z$, consistency requires that the results obtained by $W$ about the variable $S_z$ and the pointer are correlated'' (p.\ 1652, notation adapted).}\ This ensures that if these two agents interact their results will agree, provided that $O$ does not destroy the information obtained after the measurement of $S_z$ by performing another observation of a magnitude incompatible with $z-$spin. Alternatively stated, if $W$ measures the state of the complex system $s+O$, he will break the superposition in \eqref{Estar} and will found the state $|\uparrow\rangle_z\otimes|O_\uparrow\rangle$---proviso that $O$ did not performed another measurement destroying the information obtained measuring $S_z$. Then, if $W$ asks to $O$ which result they obtained, the answer will be $|\uparrow\rangle_z$, so that the descriptions of $W$ and $O$ will match. This is in virtue of the decoherence process that took place within the lab, that ``stabilized'' the measurement outcome.

\subsection{Laudisa's Criticisms to RQM}

Having discussed the relational explanation of the Wigner friend though experiment, let us now take into account the major objections against RQM contained in \cite{Laudisa:2019}. In the first place, Laudisa correctly points out that the physical situation considered in the third person scenario is identical to the Wigner friend paradox; however, he underlines that Wigner's aim was not to demonstrate the relational character of quantum theory, but rather to show a fundamental ambiguity about \emph{when} the unitary dynamics provided by the Schr\"odinger equation ``should stop holding''. 

In the second place, Laudisa interestingly claims that ``in indexing states relative to observers, RQM attempts to dissolve the inconsistency that appears to follow from the linear dynamics of the states involved in a typical QM measurement process'' (\cite{Laudisa:2019}, p.\ 222). However, he says, such a move is problematic since the sequence $\textbf{E}$ described by \eqref{E} ``overlooks the correlation'' between the possible states of the system $s$ and those of the observer $O$ that is created before a definite measurement outcome is selected by the collapse process. More precisely, Laudisa argues that the correct dynamical history of what happens in the laboratory should go first from the superposed state $a|\uparrow\rangle_z+b|\downarrow\rangle_z$ to the entangled state $a|\uparrow\rangle_z\otimes|O_\uparrow\rangle + b|\downarrow\rangle_z\otimes |O_\downarrow\rangle$, and only after, in virtue of collapse, to the state $|\uparrow\rangle_z$. However, Laudisa remarks, the first step of this dynamical process ``leads immediately to the description $\textbf{E'}$:

\begin{align*}
\left.\begin{aligned}
t_1 \longrightarrow t_2\\
(a|\uparrow\rangle_z+b|\downarrow\rangle_z)\otimes |O-ready\rangle \longrightarrow a|\uparrow\rangle_z\otimes|O_\uparrow\rangle + b|\downarrow\rangle_z\otimes |O_\downarrow\rangle
\end{aligned}\right\} \textbf{E'}
\end{align*}
\vspace{1mm}

\noindent Consequently, he concludes that it is not true that the sequences of events $\textbf{E, E'}$ are different, as Rovelli's theory entails. The sequence $\textbf{E'}$, then, ``is not a different sequence w.r.t.\ to $\textbf{E}$, but simply the same sequence under the (standard) assumption that the correlation between $O$ and $s$ is taken \emph{explicitly} into due account'' (\emph{ibid.}, p.\ 222).

Finally, and more importantly, Laudisa argues that in the third person scenario there is an element of ambiguity concerning Wigner's perspective. According to RQM, $W$ describes the complex system $s+O$ without performing any action on it, although according to \cite{Rovelli:1996} the only way to obtain information in RQM is via interaction. Thus, Laudisa argues, it is reasonable to claim that the relational theory 
\begin{quote}
inherits from standard QM the ordinary way to describe the kind of interaction that is suitable to account for measurement, namely through the tensor product coupling. But in this case, we couple $\cal{H}_W$ to $(\mathcal{H}_{S}\otimes\cal{H}_O)$ so as to obtain $\cal{H}_W\otimes\mathcal{H}_{S}\otimes\cal{H}_O$, and in turn we obtain the sequence (with the obvious interpretation for the states $|W-ready\rangle, |W_\uparrow\rangle$ and $|W_\downarrow\rangle$: 
\begin{align}
\label{P}
(a|\uparrow\rangle_z+b|\downarrow\rangle_z)\otimes|O-ready\rangle \Longrightarrow \\
a|\uparrow\rangle_z\otimes|O_\uparrow\rangle\otimes|W_\uparrow\rangle + b|\downarrow\rangle_z\otimes|O_\downarrow\rangle\otimes|W\downarrow\rangle \nonumber
\end{align}
\noindent which is completely consistent with \textbf{E} and \textbf{E'}, provided that now the system under scrutiny is $W+s+O$ and no more $s+O$. 

To sum up: the ``third person problem'' is allegedly one of the cornerstones of RQM, and the main argument underlying it is meant to support the shift to a fundamental relativization of states of quantum physical systems to observers. The formulation itself of the problem, however, seems to be based on a basic ambiguity, affecting the very description of a measurement process provided by different observers. If one tries to remove the
ambiguity by sharpening the description, the alleged difference between \textbf{E} and \textbf{E'} appears to evaporate and the motivation for the relativization gets considerably weaker (\emph{ibid.}, pp.\ 222-223, equation and notation adapted).
\end{quote}

Laudisa's objections are very well articulated and go directly to the heart of RQM discussing some potential difficulties of this interpretative framework. In fact, it is essential to explain how RQM is able to overcome these criticisms in order to show that this theory is not affected by the conceptual ambiguities mentioned above. Thus, let us consider them in order.

In the first place, it should be noted that in his works on RQM Rovelli never stated that Wigner's friend paradox was constructed to show the relational character of quantum theory. In fact, although it is true that the third person problem and Wigner's thought experiment employ the same logical structure---and thus are formally equivalent---the explanations for the very same scenario provided by these two authors pull in opposite directions. On the one hand, Wigner argued that (i) there exists a fundamental tension (or even worse an incoherence) between the dynamical laws of quantum theory, and (ii) the cut between quantum system and classical observer is ambiguous in the standard formulation of QM. As is known, he suggested to resolve these issues with the introduction of a conscious mind whose role is to cause the suppression of the unitary dynamics and, thereby, to define clearly the moment in which the wave function should collapse. In this manner, the vagueness about the demarcation between observed system and observer is resolved as well. On the other hand, Rovelli showed a direct application of the principle of relativization of states---one of the core interpretative postulates of RQM---via the third person problem. As already stated, in \cite{Rovelli:1996} such a scenario has a central explanatory and heuristic function, since it illustrates that endorsing the metaphysical assumptions of relational QM and taking the formalism of quantum theory at face value, one is led to the relativity of states. Against this background, it is relevant to underline that RQM rejects one of the implicit premises of the Wigner's friend thought experiment, namely that there exists a unique, absolute reality which is described inconsistently by agents $O$ and $W$. Hence, contrary to Wigner's conclusions, according to RQM there is no contradiction between the descriptions provided by these two different observers. Moreover, according to Rovelli there is no incoherence between the laws of QM, since $\psi$ has no ontological significance in RQM, as we have seen in Section \ref{QMP}. In addition, as already stressed, given that in RQM every physical system can play the role of observer and observed system, the issue of the arbitrary distinction between the two simply vanishes without introducing consciousness in QM. 

In the second place, contrary to Laudisa's claim, I argue that the sequences of physical events $\textbf{E}$ and $\textbf{E'}$ are different in the context of RQM. Taking into account the sequence \eqref{E}, agent $O$ measures at some arbitrary initial time $t_{1}$ the quantity $S_z$ on the system $s$, and finds at a successive final time $t_{2}$ a determinate value for the $z-$spin in virtue of the collapse process occurred in the laboratory that suppressed the superposition of states \eqref{suppos}. The obtained measurement outcome is a fact relative to $O$. From standard QM we know that before the measurement $s$ is in a superposition of states and $O$ is in a neutral state $|O-ready\rangle$, their state in fact is: $(a|\uparrow\rangle_z+b|\downarrow\rangle_z)|O-ready\rangle$. Successively, when their physical interaction begins, the state of the system and the state of $O$ get correlated ($a|\uparrow\rangle_z\otimes|O_\uparrow\rangle + b|\downarrow\rangle_z\otimes |O_\downarrow\rangle$), until at some final time $t_{2}$ the projection rule breaks the superposition and $O$ eventually obtains a definite value for the measured quantity---in our example $|\uparrow\rangle_z$---and consequently $O$ will see the pointer of the experimental apparatus pointing in the $z-$spin-up direction. 

This situation is remarkably different for observer $W$, since his description terminates at $t_2$ with the superposition of states $a|\uparrow\rangle_z\otimes|O_\uparrow\rangle + b|\downarrow\rangle_z\otimes |O_\downarrow\rangle$, where no definite eigenvalue for $S_z$ has been obtained. Laudisa's objection in my opinion overlooks the following point: although the superposition in \eqref{Estar} appears also in the sequence $\textbf{E}$, in the latter case this is just a \emph{dynamical transition} to arrive at the final state in which $O$ obtains a sharp value for the quantity $S_z$ measured on $s$. Alternatively stated, the dynamical histories that take place according to $O$ and $W$ are not both governed by a unitary evolution contrary to Laudisa's view: from the laboratory's perspective the Schr\"odinger dynamics is suppressed by a stochastic collapse of the wave function which in turns allows agent $O$ to find the system $s$ in a definite $z-$spin state, and thereby, to record a sharp value for the measured quantity $S_z$. Thus, although the two sequences have partial structural similarity, they denote different states of affairs, since the perspectives of the observers $O$ and $W$, and their information about the measure of the $z-$spin on $s$ do not agree. In fact, albeit the actualization of a particular eigenvalue of $S_z$ is a stable fact for Wigner in virtue of decoherence, this is not a fact relative to him. Hence, we can conclude that $\textbf{E}\neq\textbf{E'}$. Moreover, Laudisa is implicitly assuming that $O$ must include themselves into their description of the sequence \eqref{E}; however, this is not the necessarily case: $O$ is merely describing what happens to $s$ relative to their own perspective, without including themselves into the sequence of physical events occurring in the lab---this is why the superposition \eqref{Estar} does not appear explicitly in \textbf{E}, but it is present in \textbf{E'} where $O$ is part of the observed system. 

Finally, let us try to shed light on the alleged ambiguity present in RQM treatment of the quantum measurement process. Referring to this, it is worth noting that the sequence of events represented by \eqref{E} constitutes not only a fact relative to $O$, but also a stable fact for $W$ in virtue of the decoherence processes taking place in the lab. However, since Wigner has not yet interacted with the complex system $s+O$, he assigns a relative state to $s+O$ in which interference effects are still present. In other words, the actualization of a definite value for $s$ is not a fact relative to $W$. The distinction between relative and stable facts, hence, shows that there is no ambiguity in the description of physical events from the perspectives of $O$ and $W$. Consequently, we conclude by saying that sequences $\textbf{E}$ and $\textbf{E'}$ represent different states of affairs in RQM. 
In addition, as already pointed out a few lines above, considering the sequence \eqref{E} one notes that $O$ is \emph{not} involved in the description of the physical events occurring in the lab. This is because \eqref{E} represents a set of facts relative to $O$; furthermore, $O$ is never in a superposed state from their own perspective within the room. Only taking into account another agent $W$, $O$ becomes part of the observed system $s+O$. Applying the same logic, equation \eqref{P} is not describing Wigner's perspective, since $W$ is not himself part of his description of what happens in the lab between $s$ and $O$.\footnote{It should be noted that also in the original Wigner's friend scenario, Wigner is not part in the description of the experimental situation taking place in the lab.} Thus, Laudisa's equation \eqref{P} seems to refer to the perspective of a super-observer $Q$, who is observing the tripartite system $W-s-O$, and not to $W$'s perspective.

In sum, we can conclude that there is no fundamental ambiguity in the relational explanation of the Wigner friend thought experiment as well as in the relativization of states.

\section{Discussion}
\label{disc}

Having explained how RQM dissolves the Wigner's friend paradox as well as the QMP, one may raise the following objections: ``(i) the relational solution to these issues is not satisfactory because the theory employs physically ill-defined notions as ``observer'' and ``measurement'', and (ii) RQM does not specify what objects determine the actualization of measurement outcomes, i.e.\ it does not provide a clear ontology for the microscopic regime''.\footnote{This objection can be found e.g.\ in \cite{Laudisa:2019}, as underlined in \cite{Rovelli:2022}, p.\ 1066. In the latter essay Rovelli responds arguing that RQM rejects a strong form of realism where physical objects have well-defined values for their properties at all times. Moreover, he continues, not only in relational QM there is no such a thing as the wave function of the universe, but also ``there is no coherent global view available''. Referring to this, \cite{Laudisa:2019b} claim that if ``by realism we mean the assumption that the world is ``out there'', irrespectively of our mental states, or perceptions, there is nothing in RQM that contradicts realism. But if by realism we mean the stronger assumption that each variables of each subsystem of the world has a single value at each and every time, then this strong version of realism is weakened by RQM''.} 

These are common criticisms to RQM which deserve to be addressed and discussed. For the sake of precision, it should be noted that such objections implicitly assume that the physical description of the quantum regime must be given employing an object-oriented ontology that specifies the elementary objects of RQM---what John S.\ Bell would call the ``beables'' of the theory---and how these entities dynamically evolve in space and time. 

In order to answer the above criticisms, it is sufficient to look closely at the principles of RQM and how they works, keeping in mind that such interpretative framework implements a completely different metaphysical view w.r.t.\ local beables theories. As we already pointed out, contrary to the case of standard QM, in Rovelli's view the notion of ``observer'' is clearly defined: every physical object that can have a state of motion---i.e.\ that can interact with other systems---can be an observer. It is a logical consequence, therefore, that any physical system can be an observer as well as an observed system. A clear example of this fact is given by the different roles played by agent $O$ in the Wigner's friend scenario. Hence, in RQM any given measurement situation---more clearly, any interaction---determines without ambiguities which system is observed and which system is observing. Moreover, as already stressed, an observer is not necessarily a conscious being, as clarified by Smerlak and Rovelli:
\begin{quote}
[a]n observer, in the sense used here, does not need to be, say ``complex'', or even less so ``conscious''. An atom interacting with another atom can be considered an observer. Obviously this does not mean that one atom must be capable of storing the information about the other atom, and consciously computing the outcome of its future interaction with it; the point is simply that the history of its past interaction \emph{is} in principle sufficient information for this computation (\cite{Smerlak:2007}, p.\ 430).
\end{quote}

Taking into account the notion of measurement, RQM simply says that any physical interaction between two or more systems constitutes an observation. Thus, it is fair to claim that there is no vagueness concerning the theoretical and observational terms appearing within the principles of the theory. Nonetheless, it must be stressed again that RQM does not provide a story or a justification of the dynamical processes causing a certain actualization of a particular quantum event, i.e.\ a wave function collapse. As \cite{Laudisa:2019b} explain, ``quantum mechanics gives probabilities for quantum events to happen, not a story representing how they happen. This core aspect of quantum theory is not resolved in RQM: it is taken as a fact of the world. What RQM does resolve is the question of when this happens: any time one system affects another one, it happens relative to this other system. What RQM does, is to show that this is not in contradiction with the existence of interference effects. But the core discreteness of the quantum event actualization is not `explained' in RQM: it is understood as the picture of how nature works according to quantum theory''.\footnote{It goes without saying that for many authors this feature of RQM constitutes an inherent defect of the theory, and thus, a justified motivation to discard it. It should be noted, however, that other well-known, established interpretations of the quantum formalism involving either a branching process, or a relativization of states seem to have a descriptive limit similar to RQM. For instance, the many worlds interpretation does not explain the physical details responsible for the branching process. Analogously, Everett's interpretation requires that in measurement situations every possible outcome is actualized in relative states corresponding to different branches of the wave function. However, the details concerning the actualization of determined records in such relative states are not provided.}

Finally, let us discuss what kind of ontology is implemented by RQM. As anticipated at the outset of this section, the theory does not employ an object-oriented ontology where the macroscopic reality is somehow reduced to the dynamical evolution of a given class of objects equipped with noncontextual properties. On the contrary, RQM as originally conceived in \cite{Rovelli:1996} implements an event ontology.\footnote{This kind of ontology has a robust philosophical basis, for a general introduction the reader may refer to \cite{Casati:2020}.} According to this proposal, events are physical interactions occurring among two (or more) systems. These interactions in turn change the values of some properties instantiated by physical systems relative to other systems functioning as observers. Thus, conforming to this event based formulation of RQM, the world is just ``an evolving network of sparse relative events, described by punctual relative values of physical variables'' (\cite{Laudisa:2019b}), as we saw in section \ref{intro}. 

Interestingly, recent developments of RQM have been helpful in clarifying the ontological content of this theory and proposing alternative metaphysical readings. In addition to the already mentioned papers by Di Biagio and Rovelli, let us mention other approaches present in literature: 

\begin{itemize}

\item \cite{Candiotto:2017} contends that RQM is best appreciated in terms of an objectless metaphysics of relations without relata. In her view Rovelli's theory should be considered an instantiation of the ontology proposed by ontic structural realism according to which the fundamental building blocks of nature are relations, not objects. Underlining the centrality of relations in RQM, Candiotto affirms that in this theory there are no objects entering into a given relation: it is the latter that generates objects. Hence, she concludes that the natural framework to cast relational quantum mechanics and to understand its metaphysical content is a radical form of structural realism;

\item Contrary to Candiotto's view, \cite{Oldofredi:2020} claims the notion of physical object can be retained in RQM implementing an ontology of properties.\footnote{Such ontologies not only have a long and prestigious philosophical tradition, but also they are endorsed by several authors in the contemporary philosophy of quantum mechanics, as for instance in \cite{Lombardi:2013}, \cite{Lombardi:2016} and \cite{Kuhlmann:2010aa}.} According to this proposal, physical systems in RQM can be defined as a mereological fusion of properties, where (i) the intrinsic attributes characterizing a certain species of particles have observer-independent values (e.g.\ mass and charge), (ii) the extrinsic qualities take definite values relative to particular observers (e.g.\ position, spin along a given axis, \emph{etc.}), and (iii) not every observable defining a certain system can have a definite value due to their contextual nature. Consequently, from the perspective of mereological bundle theory, one can properly speak about material objects in motion in spacetime, whose attributes vary in relation to different observers. This interpretation entails that RQM can be made compatible with moderate structural realism;

\item In \cite{Calosi:2020} it is argued that metaphysical indeterminacy offers a new perspective on the metaphysics of RQM. On the one hand, it is shown that indeterminacy can underwrite a full-blooded realistic attitude towards non-interacting quantum systems. On the other hand, it is suggested that RQM provides example of genuine metaphysical indeterminacy. As far as non-interacting quantum systems are concerned;

\item \cite{Dorato:2016} argues instead that isolated systems in RQM can be interpreted as objects having dispositions that are manifested through interactions with other physical systems. Such manifestations are responsible for observer-dependent measurable properties;

\end{itemize}

\noindent These examples clearly suggest that RQM can be provided with a sound metaphysics and a well-defined ontology; thus, we completed our answer to the objection considered at the beginning of this section.\footnote{One may easily rephrase the relational solution to the Wigner friend paradox and the QMP in terms of the various proposals mentioned in this section; however, this exercise is left to the reader.} 

In sum, although this theory entails a radical departure from commonly accepted physical intuitions and a new perspective about reality, we can conclude that the technical and ontological issues affecting standard quantum theory find a solution in RQM.

\section{Conclusion}
\label{conc}

In this essay we discussed how RQM dissolves the QMP and how it explains the Wigner's friend paradox, answering also to interesting objections raised by Laudisa. In particular, we have seen (i) which propositions composing the ``problem of outcomes'' are at odds w.r.t.\ the metaphysical assumptions of Rovelli's theory, and (ii) that there is no fundamental tension between the Schr\"odinger dynamics and the collapse postulate in the context of RQM.
Moreover, it has been shown how the relativization of states actually works in the context of Wigner's thought experiment, stressing the relevance of the distinction between relative and stable facts for the explanation of this alleged paradox. Finally, we briefly mentioned various proposals for the interpretation of RQM, claiming that this theoretical framework can be considered a metaphysically sound formulation of quantum theory. 

Although in this paper we argued that RQM is not affected by the well-known conundra plaguing standard QM, a lot of work remains to be done in the context of this framework. Future research will include a clarification of the nature of quantum probabilities involved within this interpretation, as well as a detailed discussion about the status of macroscopic superpositions. Moreover, it will be interesting to extend RQM to the quantum theory of fields. These exciting questions will be addressed in future works.
\vspace{5mm}

\textbf{Acknowledgements:} I warmly thank the Guest Editor of this special issue Claudio Calosi for having invited me to contribute. My sincere thanks go also to the referee of this paper, whose comments and remarks improved the quality of the essay. AO is grateful to the Funda\c{c}$\tilde{\mathrm{a}}$o para a Ci\^encia e a Tecnologia (FCT) for financial support (Grant no. 2020.02858.CEECIND).

\clearpage

\bibliographystyle{apalike}
\bibliography{PhDthesis}

\begin{thebibliography}{}

\bibitem[Adlam and Rovelli, 2022]{Adlam:2022}
Adlam, E. and Rovelli, C. (2022).
\newblock {Information is Physical: Cross-Perspective Links in Relational
  Quantum Mechanics}.
\newblock {\em Manuscript available at arXiv:2203.13342v1}, pages 1 -- 22.

\bibitem[Allard~Gu{\'e}rin et~al., 2021]{Brukner:2021}
Allard~Gu{\'e}rin, P., Baumann, V., Del~Santo, F., and Brukner, C. (2021).
\newblock A no-go theorem for the persisten reality of {W}igner's friend's
  perception.
\newblock {\em Nature Communications Physics}, 4.

\bibitem[Barrett, 2018]{Barrett:2018}
Barrett, J. (2018).
\newblock {Everett's Relative-State Formulation of Quantum Mechanics}.
\newblock {\em {Stanford Encyclopedia of Philosophy}}.

\bibitem[Bassi and Ghirardi, 2003]{Bassi:2003}
Bassi, A. and Ghirardi, G.~C. (2003).
\newblock Dynamical reduction models.
\newblock {\em Physics Reports}, 379:257--426.

\bibitem[Bell, 1987]{Bell:2004aa}
Bell, J.~S. (1987).
\newblock {\em Speakable and unspeakable in quantum mechanics}.
\newblock Cambridge University Press.

\bibitem[Bohm, 1952]{Bohm:1952ab}
Bohm, D. (1952).
\newblock {A suggested interpretation of the quantum theory in terms of
  ``hidden'' variables. II}.
\newblock {\em Physical Review}, 85(2):180--193.

\bibitem[Bong et~al., 2020]{Bong:2020}
Bong, K.-W., Utreras-Alarc{\'o}n, A., Ghafari, F., Liang, Y.-C., Tischler, N.,
  Cavalcanti, E., Pryde, G., and Wiseman, H. (2020).
\newblock {A strong no-go theorem on the Wigner's friend paradox}.
\newblock {\em Nature Communications}, 16:1199--2005.

\bibitem[Bricmont, 2016]{Bricmont:2016aa}
Bricmont, J. (2016).
\newblock {\em Making sense of Quantum Mechanics}.
\newblock Springer.

\bibitem[Calosi and Mariani, 2020]{Calosi:2020}
Calosi, C. and Mariani, C. (2020).
\newblock {Quantum relational indeterminacy}.
\newblock {\em Studies in History and Philosophy of Science Part B: Studies in
  History and Philosophy of Modern Physics}, 71:158--169.

\bibitem[Candiotto, 2017]{Candiotto:2017}
Candiotto, L. (2017).
\newblock {The Reality of Relations}.
\newblock {\em {Giornale di Metafisica}}, 2:537--551.

\bibitem[Casati and Varzi, 2020]{Casati:2020}
Casati, R. and Varzi, A. (2020).
\newblock {Events}.
\newblock {\em {Stanford Encyclopedia of Philosophy}}.

\bibitem[da~Costa et~al., 2013]{Lombardi:2013}
da~Costa, N., Lombardi, O., and Lastiri, M. (2013).
\newblock {A modal ontology of properties for quantum mechanics}.
\newblock {\em Synthese}, 190:3671--3693.

\bibitem[Di~Biagio and Rovelli, 2021]{Rovelli:2021}
Di~Biagio, A. and Rovelli, C. (2021).
\newblock {Stable Facts, Relative Facts}.
\newblock {\em Foundations of Physics}, 51(30).

\bibitem[Di~Biagio and Rovelli, 2022]{Rovelli:2021a}
Di~Biagio, A. and Rovelli, C. (2022).
\newblock {Relational Quantum Mechanics is about Facts, not States: A reply to
  Pienaar and Brukner}.
\newblock {\em Foundations of Physics}, 52.

\bibitem[Dorato, 2016]{Dorato:2016}
Dorato, M. (2016).
\newblock {Rovelli's Relational Quantum Mechanics, Anti-Monism, and Quantum
  Becoming}.
\newblock In Marmodoro, A. and Yates, D., editors, {\em {The Metaphysics of
  Relations}}, pages 235--262. Oxford University Press.

\bibitem[D{\"u}rr and Teufel, 2009]{Durr:2009fk}
D{\"u}rr, D. and Teufel, S. (2009).
\newblock {\em Bohmian mechanics: the physics and mathematics of quantum
  theory}.
\newblock Berlin: Springer.

\bibitem[Everett, 1957]{Everett:1957}
Everett, H. (1957).
\newblock ``{R}elative state'' formulation of quantum mechanics.
\newblock {\em Reviews of Modern Physics}, 29(3):454--462.
\newblock Reprinted in DeWitt, B.\ S.\ and Graham, N., editors (1973).
  \emph{The Many-Worlds Interpretation of Quantum Mechanics}, pages 141--149.
  Princeton: Princeton University Press.

\bibitem[Frauchiger and Renner, 2018]{Frauchiger:2018}
Frauchiger, D. and Renner, R. (2018).
\newblock {Quantum Theory Cannot Consistently Describe the Use of Itself}.
\newblock {\em Nature Communications}, 9:1--10.

\bibitem[Ghirardi, 2016]{Ghirardi:2016}
Ghirardi, G.~C. (2016).
\newblock {Collapse Theories}.
\newblock {\em {Stanford Encyclopedia of Philosophy}}.

\bibitem[Holland, 1993]{Holland:1993}
Holland, P. (1993).
\newblock {\em {The Quantum Theory of Motion. An Account of the de Broglie-Bohm
  Causal Interpretation of Quantum Mechanics}}.
\newblock Cambridge: Cambridge University Press.

\bibitem[Kuhlmann, 2010]{Kuhlmann:2010aa}
Kuhlmann, M. (2010).
\newblock {\em The ultimate constituents of the material world. In search of an
  ontology for fundamental physics}.
\newblock Frankfurt (Main): Ontos.

\bibitem[Laudisa, 2019]{Laudisa:2019}
Laudisa, F. (2019).
\newblock {Open Problems in Relational Quantum Mechanics}.
\newblock {\em Journal for General Philosophy of Science}, 50:215--230.

\bibitem[Laudisa and Rovelli, 2019]{Laudisa:2019b}
Laudisa, F. and Rovelli, C. (2019).
\newblock {Relational Quantum Mechanics}.
\newblock {\em Stanford Encyclopedia of Philosophy}.

\bibitem[Lazarovici and Hubert, 2019]{Lazarovici:2019}
Lazarovici, D. and Hubert, M. (2019).
\newblock {How Quantum Mechanics Can Consistently Describe the Use of Itself}.
\newblock {\em Scientific Reports}, 470(9):1--8.

\bibitem[Lombardi and Dieks, 2016]{Lombardi:2016}
Lombardi, O. and Dieks, D. (2016).
\newblock {Particles in a Quantum Ontology of Properties}.
\newblock In Bigaj, T. and W{\"u}thrich, C., editors, {\em {Metaphysics in
  Contemporary Physics}}, pages 123 -- 144. Brill Rodopi.

\bibitem[Martin-Dussaud et~al., 2019]{Rovelli:2019}
Martin-Dussaud, P., Rovelli, C., and Zalamea, F. (2019).
\newblock {The Notion of Locality in Relational Quantum Mechanics}.
\newblock {\em Foundations of Physics}, 49:96--106.

\bibitem[Maudlin, 1995]{Maudlin:1995aa}
Maudlin, T. (1995).
\newblock Three measurement problems.
\newblock {\em Topoi}, 14:7--15.

\bibitem[Oldofredi, 2020]{Oldofredi:2020}
Oldofredi, A. (2020).
\newblock {The Bundle Theory Approach to Relational Quantum Mechanics}.
\newblock {\em Foundations of Physics}, 51(1):1--22.

\bibitem[Pienaar, 2019]{Pienaar:2019}
Pienaar, J. (2019).
\newblock {Comment on ``The Notion of Locality in Relational Quantum
  Mechanics''}.
\newblock {\em Foundations of Physics}, 49:1404--1414.

\bibitem[Pienaar, 2021a]{Pienaar:2021}
Pienaar, J. (2021a).
\newblock {A Quintet of Quandaries: FIve No-Go Theorems for Relational Quantum
  Mechanics}.
\newblock {\em Foundations of Physics}, 51(97).

\bibitem[Pienaar, 2021b]{Pienaar:2021a}
Pienaar, J. (2021b).
\newblock {QBsim and Relational Quantum Mechanics compared}.
\newblock {\em Foundations of Physics}, 51.

\bibitem[Rovelli, 1996]{Rovelli:1996}
Rovelli, C. (1996).
\newblock {Relational Quantum Mechanics}.
\newblock {\em International Journal of Theoretical Physics}, 35(8):1637--1678.

\bibitem[Rovelli, 2016]{Rovelli:2016}
Rovelli, C. (2016).
\newblock {An Argument Against the Realistic Interpretation of the Wave
  Function}.
\newblock {\em Foundations of Physics}, 46:1229 -- 1237.

\bibitem[Rovelli, 2018]{Rovelli:2018}
Rovelli, C. (2018).
\newblock {Space is blue and birds fly through it}.
\newblock {\em Philosophical Transactions of the Royal Society A, Physical and
  Engineering Sciences}, 376(2123):2017.0312.

\bibitem[Rovelli, 2022]{Rovelli:2022}
Rovelli, C. (2022).
\newblock {The Relational Interpretation}.
\newblock In Freire, O.~J., editor, {\em {The Oxford Handbook of the History of
  Quantum Interpretations}}, chapter~43, pages 1055--1071. Oxford University
  Press.

\bibitem[Rovelli and Vidotto, 2014]{Rovelli:2014}
Rovelli, C. and Vidotto, F. (2014).
\newblock {\em {Covariant Loop Quantum Gravity: An Elementary Introduction to
  Quantum Gravity and Spinfoam Theory}}.
\newblock Cambridge University Press.

\bibitem[Smerlak and Rovelli, 2007]{Smerlak:2007}
Smerlak, M. and Rovelli, C. (2007).
\newblock {Relational EPR}.
\newblock {\em Foundations of Physics}, 37:427--445.

\bibitem[Wallace, 2012]{Wallace:2012aa}
Wallace, D. (2012).
\newblock {\em The emergent multiverse. Quantum theory according to the
  {E}verett interpretation}.
\newblock Oxford: Oxford University Press.

\bibitem[Wigner, 1995]{Wigner:1995}
Wigner, E. (1995).
\newblock {Remarks on the mind-body question}.
\newblock In Mehra, J., editor, {\em {Philosophical Reflections and Syntheses.
  The Collected Works of Eugene Paul Wigner (Part B Historical, Philosophical,
  and Socio-Political Papers), vol B / 6.}} Springer.

\end{thebibliography}
\end{document}